\documentstyle[12pt]{article}

\begin{document}

\begin{center}
Hydrodynamical Approach to Vehicular Flow in the Urban Street Canyon
\end{center}
\begin{center}
Maciej M. Duras
\end{center}
\begin{center}
Institute of Physics, Cracow University of Technology, 
ulica Podchor\c{a}\.zych 1, PL-30084 Cracow, Poland
\end{center}

\begin{center}
Email: mduras @ riad.usk.pk.edu.pl
\end{center}

\begin{center}
AD 2001 March 22
\end{center}

\begin{center}
"DCOMP01, The 2001 Annual Meeting of the Division of Computational Physics"; 
June 25th, 2001 to June 28th, 2001; 
Massachusetts Institute of Technology, Cambridge, Massachusetts, USA; 
in:  "DCOMP Meeting 2001 Scientific Program DCOMP June 25-28, 2001; 
Cambridge, Massachusetts"; {\em Bulletin of the American Physical Society},   
Vol. {\bf 46}, Part {\bf F}, R1.030 (2001).
\end{center}

\begin{abstract}
The vehicular flow in the urban street canyon is considered. 
The classical field description is used in the modelling of 
the vehicular movement and of gaseous mixture in generic 
urban street canyon. The dynamical variables include 
vehicular densities, velocities, and emissivities: of 
pollutants, heat and exhaust gases, as well as standard 
mixture components' variables: densities, velocities, 
temperature, pressures. The local balances' equations 
predict the dynamics of the complex system. The automatic 
control of the vehicular flow is attained by the sets of 
coordinated traffic lights. The automatic control is aimed 
at minimization of traffic ecological costs by the 
application of variational calculus (Lagrange's and Bolz's 
problems). The theoretical description is accompanied by 
numerical examples of computer fluid dynamics based on real 
traffic data.
\end{abstract}
\section{Description of the model.}
\label{sect-description}
\setcounter{equation}{0}  
In the present article we develop a continuum field model of the street canyon.
In the next article we will deal with numerical examples 
\cite{Duras 1999 engtrans numer}.
The vehicular flow in the canyon is multilane bidirectional one-level
rectilinear, and it is considered
with two coordinated signalized junctions
\cite{Duras 1998 thesis, Duras 1997 PJES, Duras 1999 PJES}.
The vehicles belong to different vehicular classes:
passenger cars, and trucks.
Emissions from the vehicles are based on technical measurements
and many types of pollutants are considered
(carbon monoxide CO, hydrocarbons HC, nitrogen oxides ${\rm NO_{x}}$).
The vehicular dynamics is based on a hydrodynamical approach
\cite{Michalopoulos 1984}.
The governing equations are the continuity equation
for the number of vehicles, 
and Greenshields' equilibrium speed-density u-k model
\cite{Greenshields 1934}.

The model of dynamics of pollutants is also hydrodynamical.
The model consists of a set of mutually interconnected
nonlinear, three-dimensional, time-dependent, 
partial differential equations with nonzero right-hand sides (sources),
and of boundary and of initial problem.
The pollutants, oxygen, and the remaining gaseous constituents of air,
are treated as mixture of noninteracting, Newtonian, viscous
fluid (perfect or ideal gases).
The complete model incorporates as variables the following fields:
density of the mixture, mass concentrations of constituents of the mixture,
velocity of mixture, temperature of mixture, pressure of mixture,
intrinsic (internal) energy of mixture,
densities of vehicles, and velocities of vehicles.
The model is based on the assumption 
of local laws of balance (conservation) 
of: mass of the mixture, masses of its constituents,
momentum and energy of the mixture,
the numbers of the vehicles,
as well as of the state equations (Clapeyron's law and Greenshields' model).
The equations of dynamics are solved by the finite difference
scheme.

The six separate monocriterial optimization problems are formulated
by defining the functionals of total travel time,
of global emissions of pollutants, 
and of global concentrations of pollutants,
both in the studied street canyon
and in its two nearest neighbour substitute canyons.
The vector of control is a five-tuple
composed of two cycle times, two green times,
and one offset time between the traffic lights.
The optimal control problem consists of
minimization of the six functionals
over the admissible control domain.

\section{Equations of dynamics.}
\label{sect-dynamics}
\setcounter{equation}{0}  
Under the above model specifications, 
the  complete set of equations of dynamics of the model 
is formulated as follows
(we follow the general idea presented  
in \cite{Eringen 1990, Landau 1986}):

{\bf E1. Balance of momentum of mixture - Navier Stokes equation.}
\begin{equation}
\rho (\frac{\partial {\bf{v}}}{\partial t}
+ ({\bf{v}} \circ \nabla) {\bf{v}}) + S{\bf{v}}=
- \nabla p + \eta \Delta {\bf{v}} 
+ (\xi + \frac{\eta}{3}) \nabla ({\rm div} {\bf{v}}) + {\bf{F}},
\label{Navier-Stokes-eq}
\end{equation}
where $\eta$ is the first viscosity coefficient 
($\eta=18.1 \cdot 10^{-6} [\frac{{\rm kg}}{{\rm m \cdot s}}]$ 
for air at temperature $T=293.16$ [K]), 
$\xi$ is the second viscosity coefficient 
($\xi=15.6 \cdot 10^{-6} [\frac{{\rm kg}}{{\rm m \cdot s}}]$ 
for air at temperature $T=293.16$[K]), 
${\bf{F}}=\rho {\bf{g}}$ is the gravitational body force density,
${\bf{g}}$ is the gravitational acceleration 
of Earth (${\bf{g}}=(0, 0, -9.81) [\frac{{\rm m}}{{\rm s^{2}}}]$),
$\nabla \bf{v}$ is gradient of the vector (so it is a tensor of rank 2). 
We assume that the gaseous mixture is a compressible and 
viscous fluid.

{\bf E2. Balance of mass of mixture - Equation of continuity.}
\begin{equation}
\frac{\partial \rho}{\partial t}
+ {\rm div}(\rho {\bf{v}})=S.
\label{continuity-eq}
\end{equation}  

We have assumed the source {\bf D1}.

{\bf E3. Balances of masses of constituents of mixture - Diffusion equations.}
 
{\bf E3a.}
\begin{eqnarray}
& & \rho (\frac{\partial c_{i}}{\partial t} + {\bf{v}} \circ \nabla c_{i})=
S^{E}_{i} - c_{i} S + 
\label{diffusion-eq-a} \\
& & + \sum_{m=1}^{N-1}\{ (D_{im}-D_{iN})
\cdot {\rm div}[\rho \nabla (c_{m}+\frac{k_{T, m}}{T} \nabla T)] \},
i=1, ..., N_{E}. \nonumber
\end{eqnarray}

{\bf E3b.}
\begin{eqnarray}
& & \rho (\frac{\partial c_{i}}{\partial t} + {\bf{v}} \circ \nabla c_{i})=
- c_{i} S +
\label{diffusion-eq-b} \\
& & + \sum_{m=1}^{N-1}\{ (D_{im}-D_{iN})
\cdot {\rm div}[\rho \nabla (c_{m}+\frac{k_{T, m}}{T} \nabla T)] \},
i=(N_{E}+1), ..., N, \nonumber
\end{eqnarray}
where $D_{im}=D_{mi}$ is the mutual diffusivity coefficient 
from the $i$-th constituent to $m$-th one, and $D_{ii}$  is the 
autodiffusivity coefficient of the $i$-th constituent,
and $k_{T, m}$ is the thermodiffusion ratio of the $m$-th constituent. 
The diffusivity coefficients and thermodiffusion ratios 
are constant and known).
In {\bf E3a} we have assumed the sources {\bf D1-D2}. 
In {\bf E3b} only the source {\bf D1} is taken into account. 
Since the mixture is in motion, we cannot neglect the convection term:
${\bf{v}} \circ \nabla c_{i}$.  
We assume that the barodiffusion and gravitodiffusion coefficients 
are equal to zero.

{\bf E4. Balance of energy of mixture.}
\begin{equation}
\rho (\frac{\partial \epsilon}{\partial t} + {\bf{v}} \circ \nabla \epsilon)=
-(-\frac{1}{2}{\bf{v}}^{2} + \epsilon) S
+ {\bf T}:\nabla {\bf{v}}+{\rm div}(-{\bf{q}})+\sigma,
\label{energy-eq} 
\end{equation}
where $\epsilon$ is the mass density of intrinsic (internal) energy 
of the air mixture, 
${\bf T}$ is the stress tensor, 
symbol $:$ denotes the contraction operation,
$\bf{q}$ is the vector of flux of heat.
We assume that \cite{Duras 1998 thesis}:
\begin{eqnarray} 
& & \epsilon=\sum_{i=1}^{N}\epsilon_{i},
\label{intrinsic-energy-def} \\
& & \epsilon_{i}=
\frac{1}{m_{i}} 
\{ 
c_{i} k_{B} T 
\exp(-\frac{m_{i}|{\bf{g}}|z}{k_{B}T}) \cdot
[
(-\frac{z}{c}) \cdot (1-\exp(-\frac{m_{i}|{\bf{g}}|c}{k_{B}T}))
-\exp(-\frac{m_{i}|{\bf{g}}|c}{k_{B}T})
]
\} +
\label{ith-intrinsic-energy-def} \\
& & + \tilde{\mu}_{i}c_{i}, 
\nonumber \\
& &
\tilde{\mu}_{i}=\frac{\mu_{i}}{m_{i}},
\label{chemical-potential-tilde} \\
& & 
\mu_{i}= 
k_{B} T \cdot
\{
\ln
[
(c_{i} p) (k_{B} T)^{-\frac{c_{p, i}}{k_{B}}}
(\frac{m_{{\rm air}}}{m_{i}}) (\frac{2\pi h^{2}}{m_{i}})^{\frac{3}{2}}
]
\}
+m_{i} |{\bf{g}}| z
,
\label{chemical-potential-def} \\
& & {\rm T}_{mk}=-p\delta_{mk}
+
\label{stress-tensor-def} \\
& & + \eta \cdot 
[
(
\frac{\partial v_{m}}{\partial x_{k}}
+
\frac{\partial v_{k}}{\partial x_{m}}
-
\frac{2}{3} \delta_{mk} {\rm div}({\bf{v}})
)
\frac{\partial v_{k}}{\partial x_{m}}
]
+
\xi \cdot
[
(
\delta_{mk} {\rm div}({\bf{v}})
)^{2}
],
m, k=1, ...,3,
\nonumber \\
& & {\bf T}:\nabla {\bf{v}}=
\sum_{m=1}^{3} \sum_{k=1}^{3} 
{\rm T}_{mk} \frac{\partial v_{m}}{\partial x_{k}}, 
\label{stress-tensor-velocity-contraction} \\
& & {\bf{q}}=
\sum_{i=1}^{N}
\{
[
(\frac{\beta_{i}T}{\alpha_{ii}}
+ \tilde{\mu}_{i}) {\bf{j}}_{i}
]
+
[
(-\kappa) \nabla T
]
\}
,
\label{vector-flux-heat-def} \\
& & {\bf{j}}_{i}=
-\rho D_{ii} (c_{i}+\frac{k_{T, i}}{T} \nabla T)
,
\label{vector-flux-mass-def} \\
& & \alpha_{ii}=
\frac{
[\rho D_{ii}]
}
{ 
[(\frac{\partial \tilde{\mu}_{i}}{\partial c_{i}})
_{(c_{n})_{n=1, ..., N, i \neq n}, T, p}]
},
\label{alpha-def} \\
& & \beta_{i}=
[\rho D_{ii}] \cdot
\{
\frac{k_{T, i}}{T}
-
\frac{
[(\frac{\partial \tilde{\mu}_{i}}{\partial T})
_{(c_{n})_{n=1, ..., N}, p}]
}
{
[(\frac{\partial \tilde{\mu}_{i}}{\partial c_{i}})
_{(c_{n})_{n=1, ..., N, i \neq n}, T, p}]
}
\}
,
\label{beta-def}
\end{eqnarray}
where 
$\epsilon_{i}$ is the mass density of intrinsic (internal) energy 
of the $i$th constituent of the air mixture,
$m_{i}$ the molecular mass of the $i$th constituent,
$k_{B}=1.3807 \cdot 10^{-23} [\frac{{\rm J}}{{\rm kg}}]$ 
is Boltzmann's constant,
$\mu_{i}$ is the complete partial chemical potential of the $i$th constituent 
of the air mixture (it is complete since it is composed of
chemical potential without external force 
field and of external potential),
$m_{{\rm air}}=28.966$ [u] is the molecular mass of air
($1[{\rm u}]=1.66054 \cdot 10^{-27}$ [kg]),
$\delta_{mk}$ is Kronecker's delta,
$c_{p, i}$ is the specific heat at constant pressure of the $i$th 
constituent of air mixture,
$h=6.62608 \cdot 10^{-34}$ [$J \cdot s$] is Planck's constant,
${\bf{j}}_{i}$ is the vector of flux of mass of the $i$th constituent
of the air mixture,
and $\kappa$ is the coefficient of thermal conductivity of air.
These magnitudes are derived from
Grand Canonical ensemble with external gravitational Newtonian field.

{\bf E5. Equation of state of the mixture - Constitutive equation -
Clapeyron's equation.}

\begin{equation}
\frac{p}{\rho}=\frac{R}{m_{{\rm air}}} \cdot T
\label{state-eq}
\end{equation}
is Clapeyron's equation of state for a gaseous mixture,
where $R=8.3145$ [$\frac{{\rm J}}{{\rm mole \cdot K}}$] is the gas constant.
\begin{equation}
p_{i}=c_{i} \cdot \frac{m_{{\rm air}}}{m_{i}} \cdot p
\label{constituent-state-eq}
\end{equation}
are partial pressures of constituents according to Dalton's law.

{\bf E6. Balances of numbers of vehicles - Equations of continuity of vehicles.}

\begin{equation}
\frac{\partial k_{l, vt}^{s}}{\partial t} 
+ {\rm div}(k_{l, vt}^{s}{\bf{w}}_{l, vt}^{s})=0.  
\label{vehicle-continuity-eq-a}
\end{equation}  

{\bf E7. Equations of state of vehicles - Greenshields model.}

\begin{equation}
{\bf{w}}_{l, vt}^{s}(x, t)=
(
w_{l, vt, f}^{s} \cdot
(1-\frac{k_{l, vt}^{s}(x, t)}{k_{l, vt, {\rm jam}}^{s}}), 0, 0).
\label{Greenshields-eq-a} 
\end{equation}

The Greenshields equilibrium speed-density u-k model is assumed 
\cite{Greenshields 1934}. 
The values of maximum free flow speed 
$w_{l, vt, f}^{s}$, 
and of jam vehicular densities 
$k_{l, vt, {\rm jam}}^{s}$,
are given in Tables 3 and 8 of \cite{Duras 1999 engtrans numer}.
 
{\bf E8. Technical parameters.}

The dependence of emissivity on the density and velocity 
of vehicles is assumed in the form:

{\bf E8a.}

\begin{equation}
e_{l, ct, vt}^{s}(x, t)=
k_{l, vt}^{s}(x, t) \cdot
[
(\frac{|{\bf{w}}_{l, vt}^{s}(x, t)|-\tilde{w}_{ct, vt, i_{l}}}
{\tilde{w}_{ct, vt, i_{l}+1}-\tilde{w}_{ct, vt, i_{l}}}) \cdot
(\tilde{e}_{ct, vt, i_{l}+1}-\tilde{e}_{ct, vt, i_{l}})
+ 
\tilde{e}_{ct, vt, i_{l}}
], 
\label{technical-emission-eq-L} 
\end{equation}
where $\tilde{w}_{ct, vt, i_{l}}$  
are experimental velocities,
$|{\bf{w}}_{l, vt}^{s}(x, t)|
\in (\tilde{w}_{ct, vt, i_{l}}, \tilde{w}_{ct, vt, i_{l}+1}),$
$\tilde{e}_{ct, vt, i_{l}},$
are experimental 
emissions of the $ct$th exhaust gas from single vehicle 
of $vt$th type at velocities $\tilde{w}_{ct, vt, i_{l}}$,
respectively,  
measured in [$\frac{{\rm kg}}{{\rm veh \cdot s}}$],
$i_{l}=1, ..., N_{EM},$
$N_{EM}$ is the number of experimental measurements.
Similarly, the dependence of the change 
of the linear density of energy on the density and 
velocity of vehicles is taken in the form:

{\bf E8b.}
\begin{equation}
\sigma_{l, vt}^{s}(x, t)=
q_{vt} \cdot k_{l, vt}^{s}(x, t) \cdot
[
(\frac{|{\bf{w}}_{l, vt}^{s}(x, t)|-\bar{w}_{vt, i_{l}}}
{\bar{w}_{vt, i_{l}+1}-\bar{w}_{vt, i_{l}}}) \cdot 
(\sigma_{vt, i_{l}+1}-\sigma_{vt, i_{l}})
+ 
\sigma_{vt, i_{l}}
],
\label{technical-heat-eq-L} 
\end{equation} 
where 
$\sigma_{vt, i_{l}},$
are experimental values of consumption of gasoline/diesel for a single 
vehicle of $vt$th type at velocities $\bar{w}_{vt, i_{l}},$
respectively, 
measured in [$\frac{{\rm kg}}{{\rm veh \cdot s}}$],
$q_{vt}$ is the emitted combustion 
energy per unit mass of gasoline/diesel
[$\frac{{\rm J}}{{\rm kg}}$].

\section{\bf Optimization problems.}
\label{sect-optimization}
\setcounter{equation}{0}  
Our control task is the minimization of the measures 
of the total travel time (TTT) 
\cite{Michalopoulos 1984},
emissions (E), and concentrations (C) 
of exhaust gases in the street canyon, 
therefore the appropriate optimization problems 
may be formulated as follows \cite{Duras 1998 thesis}:

{\bf F0. Vector of control.}
\begin{equation}
{\bf {u}}=(g_{1}, C_{1}, g_{2}, C_{2}, F) \in U^{{\rm adm}},
\label{control-5-tuple-eq}
\end{equation}
where ${\bf {u}}$ is vector of boundary control, $g_{m}$ are green times, 
$C_{m}$ are cycle times,
$F$ is offset time, 
and $U^{{\rm adm}}$ is a set of admissible control variables
(compare {\bf A8, A9, B5, B5S, B5SS}). 

We define six functionals {\bf F1-F6} of the total travel time,
emissions, and concentrations of pollutants
in single canyon, and in canyon with the nearest neighbour substitute canyons,
respectively.

{\bf F1. Total travel time for a single canyon.} 
\begin{equation}
J_{{\rm TTT}}({\bf{u}}) 
=\sum_{s=1}^{2} \sum_{l=1}^{n_{s}} \sum_{vt=1}^{VT}
\int_{0}^{a} \int_{0}^{T_{S}}
k_{l, vt}^{s}(x, t) dx \, dt .
\label{functional-TTT-def} 
\end{equation}

{\bf F2. Global emission for a single canyon.}
\begin{equation}
J_{{\rm E}}({\bf{u}})
=\sum_{s=1}^{2} \sum_{l=1}^{n_{s}} \sum_{ct=1}^{CT} \sum_{vt=1}^{VT}
\int_{0}^{a} \int_{0}^{T_{S}}
e_{l, ct, vt}^{s}(x, t) dx \, dt .
\label{functional-E-def} 
\end{equation}

{\bf F3. Global pollutants concentration for a single canyon.}
\begin{equation}
J_{{\rm C}}({\bf{u}})=
\rho_{{\rm STP}} \cdot 
\sum_{i=1}^{N_{E}-1} 
\int_{0}^{a} \int_{0}^{b} \int_{0}^{c} \int_{0}^{T_{S}}
c_{i}(x, y, z, t) dx \, dy \, dz \, dt.
\label{functional-C-def}
\end{equation}

{\bf F4. Total travel time for the canyon in street subnetwork.}
\begin{eqnarray}
& & J_{{\rm TTT, ext}}({\bf{u}})=
J_{{\rm TTT}}({\bf{u}}) + 
\label{functional-TTT-ext-def} \\
& & +
a \cdot \sum_{s=1}^{2} \alpha_{{\rm TTT, ext}}^{s}
\sum_{l=1}^{n_{s}} \sum_{vt=1}^{VT} k_{l, vt, {\rm jam}}^{s}
\cdot (C_{s}-g_{s}).
\nonumber
\end{eqnarray}

{\bf F5. Global emission for the canyon in street subnetwork.}
\begin{eqnarray}
& & J_{{\rm E, ext}}({\bf{u}})=
J_{{\rm E}}({\bf{u}}) +
\label{functional-E-ext-def} \\
& & +
a \cdot \sum_{s=1}^{2} \alpha_{{\rm E, ext}}^{s}
\sum_{l=1}^{n_{s}} \sum_{ct=1}^{CT} \sum_{vt=1}^{VT} 
e_{l, ct, vt, {\rm jam}}^{s}
\cdot (C_{s}-g_{s}).
\nonumber 
\end{eqnarray}

{\bf F6. Global pollutants concentration for the canyon in street subnetwork.}
\begin{eqnarray}
& & J_{{\rm C, ext}}({\bf{u}})=
J_{{\rm C}}({\bf{u}}) +
\label{functional-C-ext-def} \\
& & +
\rho_{{\rm STP}} \cdot a \cdot b \cdot c \cdot 
\sum_{i=1}^{N_{E}-1} c_{i, {\rm STP}} \cdot 
\sum_{s=1}^{2} \alpha_{{\rm C, ext}}^{s} \cdot (C_{s}-g_{s}).
\nonumber
\end{eqnarray}

The integrands $k_{l, vt}^{s}, 
e_{l, ct, vt}^{s}, c_{i}$ 
in functionals {\bf F1-F6} depend on the control vector ${\bf u}$ {\bf F0}
through the boundary conditions {\bf B0-B8},
through the equations of dynamics {\bf E1-E8},
as well as, through the sources {\bf D0-D2}.
The value of the vector of control $u$ directly
affects the boundary conditions {\bf B5, B5S, B5SS},
and then the boundary conditions {\bf B6-B8}
for vehicular densities, velocities, and emissivities.
It also affects the sources {\bf D0-D2}.
Next, it propagates to the equations of dynamics {\bf E1-E8}
and then it influences the values of functionals {\bf F1-F6}.
We only deal with six monocriterial optimization problems {\bf O1-O6},
and not with one multicriterial problem.
We put the scaling parameters equal to unity:
$\alpha_{{\rm TTT, ext}}^{s}=
\alpha_{{\rm E, ext}}^{s}=
\alpha_{{\rm C, ext}}^{s}=1.0,$  
in functionals {\bf F4-F6}.
$\rho_{{\rm STP}}$ is the density of air 
at standard temperature and pressure STP,
$c_{i, {\rm STP}}$ is concentration of the $i$th
constituent of air at standard temperature and pressure.
$J_{{\rm TTT}}$ and $J_{{\rm TTT, ext}}$ are measured in [$veh \cdot s$],
$J_{{\rm E}}$ and $J_{{\rm E, ext}}$ are measured in [kg],
and $J_{{\rm C}}$ and $J_{{\rm C, ext}}$ are measured in [$kg \cdot s$],
respectively. 

Now we formulate six separate monocriterial optimization
problems {\bf O1-O6} that consist in minimization
of functionals {\bf F1-F6} with respect to control vector
{\bf F0} over admissible domain, while the equations
of dynamics {\bf E1-E8} are fulfilled.  

{\bf O1. Minimization of total travel time for a single canyon.} 
\begin{equation}
J_{{\rm TTT}}^{\star}=J_{{\rm TTT}}({\bf{u}}_{{\rm TTT}}^{\star})=\min \{ {\bf{u}}\in U^{{\rm adm}}: 
J_{{\rm TTT}}({\bf{u}}) \};
\label{optimum-TTT-def}
\end{equation}

{\bf O2. Minimization of global emission for a single canyon.}
\begin{equation}
J_{{\rm E}}^{\star}=J_{{\rm E}}({\bf{u}}_{{\rm E}}^{\star})=\min \{ {\bf{u}}\in U^{{\rm adm}}: 
J_{{\rm E}}({\bf{u}}) \};
\label{optimum-E-def}
\end{equation}   

{\bf O3. Minimization of global pollutants concentration for a single canyon.}
\begin{equation}
J_{{\rm C}}^{\star}=J_{{\rm C}}({\bf{u}}_{{\rm C}}^{\star})=\min \{ {\bf{u}}\in U^{{\rm adm}}:
J_{{\rm C}}({\bf{u}}) \};
\label{optimum-C-def}
\end{equation}   

{\bf O4. Minimization of total travel time for a canyon in street subnetwork.}
\begin{equation}
J_{{\rm TTT, ext}}^{\star}=J_{{\rm TTT, ext}}({\bf{u}}_{{\rm TTT, ext}}^{\star})
=\min \{ {\bf{u}}\in U^{{\rm adm}}: J_{{\rm TTT, ext}}({\bf{u}}) \};
\label{optimum-TTT-ext-def}
\end{equation}   

{\bf O5. Minimization of global emission for a canyon in street subnetwork.}
\begin{equation}
J_{{\rm E, ext}}^{\star}=J_{{\rm E, ext}}({\bf{u}}_{{\rm E, ext}}^{\star})
=\min \{ {\bf{u}}\in U^{{\rm adm}}: J_{{\rm E, ext}}({\bf{u}}) \};
\label{optimum-E-ext-def}
\end{equation}   

{\bf O6. Minimization of global pollutants concentration for a canyon in street subnetwork.}
\begin{equation}
J_{{\rm C, ext}}^{\star}=J_{{\rm C, ext}}({\bf{u}}_{{\rm C, ext}}^{\star})
=\min \{ {\bf{u}}\in U^{{\rm adm}}: J_{{\rm C, ext}}({\bf{u}}) \},
\label{optimum-C-ext-def}
\end{equation}   
where $J_{{\rm TTT}}^{\star}, J_{{\rm E}}^{\star}, J_{{\rm C}}^{\star}, 
J_{{\rm TTT, ext}}^{\star}, J_{{\rm E, ext}}^{\star}, J_{{\rm C, ext}}^{\star}$,
are the minimal values of the functionals {\bf F1-F6},
and ${\bf{u}}_{{\rm TTT}}^{\star}, {\bf{u}}_{{\rm E}}^{\star}, {\bf{u}}_{{\rm C}}^{\star}, 
{\bf{u}}_{{\rm TTT, ext}}^{\star}, {\bf{u}}_{{\rm E, ext}}^{\star}, {\bf{u}}_{{\rm C, ext}}^{\star}$,
are control vectors at which the functionals reach the minima, respectively.

\section{Conclusions.}
\label{sect-conclusions}
\setcounter{equation}{0}  

The proecological traffic control idea 
and advanced model of the street canyon have been developed. 
It has been found that the proposed model
represents the main features of complex air pollution phenomena.

\end{document}